\documentclass{article}
\usepackage[utf8]{inputenc}
\usepackage[margin=0.75in]{geometry}
\usepackage{mathptmx}
\usepackage[T1]{fontenc}
\usepackage{amsmath,amssymb}
\usepackage{float}
\usepackage{graphicx}
\usepackage{wrapfig}
\usepackage{xcolor}
\usepackage{authblk}
\usepackage[small,compact]{titlesec}
\usepackage[numbers,sort&compress]{natbib}
\usepackage{hyperref}
\usepackage{caption}
\usepackage{ulem}
\usepackage[left, mathlines]{lineno}
\titleformat*{\section}{\normalfont\normalsize\bfseries}

\title{Currents Beneath Stability: A Stochastic Framework for Exchange Rate Instability Using Kramers--Moyal Expansion}

\author[1]{Yazdan Babazadeh Maghsoodlo}
\author[2]{Amin Safaeesirat}

\affil[1]{\small{Department of Applied Mathematics, University of Waterloo, Waterloo, ON, Canada}}
\affil[2]{\small{Department of Physics, Simon Fraser University, Burnaby, Canada}}
\date{April 2025}

\begin{document}

\maketitle

\begin{abstract}

Understanding the stochastic behavior of currency exchange rates is critical for assessing financial stability and anticipating market transitions. In this study, we investigate the empirical dynamics of the USD exchange rate in three economies, including Iran, Turkey, and Sri Lanka, through the lens of the Kramers–Moyal expansion and Fokker–Planck formalism. Using log-return data, we confirm the Markovian nature of the exchange rate fluctuations, enabling us to model the system with a second-order Fokker–Planck equation. The inferred Langevin coefficients reveal a stabilizing linear drift and a nonlinear, return-dependent diffusion term, reflecting both regulatory effects and underlying volatility. A rolling-window estimation of these coefficients, paired with structural breakpoint detection, uncovers regime shifts that align with major political and economic events, offering insight into the hidden dynamics of currency instability. This framework provides a robust foundation for detecting latent transitions and modeling risk in complex financial systems.
\end{abstract}
\section{Introduction}

The stochastic nature of currency exchange rates plays a fundamental role in the dynamics of global financial markets. These rates are influenced by a complex interplay of factors, including interest rate differentials, inflation expectations, speculative trading, news \cite{laakkonen2004impact}, and geopolitical developments \cite{yilmazkuday2025geopolitical}, which introduce inherent randomness into their fluctuations \cite{ramasamy2015influence}. Understanding these stochastic behaviors is not only an academic exercise, but it is also critical for a wide range of practical applications. For policymakers, accurate models of exchange rate volatility inform the design of monetary interventions, help stabilize inflation expectations, and protect foreign reserves. For investors and risk managers, such insights are vital for hedging strategies and forecasting currency risk exposures. In addition, in countries experiencing increased political uncertainty, sanctions, or structural economic fragility, exchange rates can become indicators of deeper systemic instabilities. As a result, modeling these fluctuations with robust stochastic frameworks is essential for detecting early signs of financial instability and anticipating potential crises \cite{albulescu2010forecasting}.

Recent studies have increasingly applied stochastic differential equations (SDEs) in a wide range of financial modeling problems. Originally central to classical models like Black-Scholes and Hull-White, SDEs have been significantly expanded to capture complex dynamics such as wealth evolution \cite{osok2025modelling}, jumps \cite{chirima2024pricing}, regime switches \cite{pelizzari2025topics}, mortgage pricing \cite{jalili2025new}, and memory effects \cite{emmanuelhull,renner2001evidence}. Newer approaches, such as forward-backward SDEs demonstrated in deep learning frameworks \cite{ashok2024stable}, show promise for robust option pricing and valuation adjustments calculations. These works demonstrate that SDEs now serve not only as theoretical instruments but also as computationally efficient tools for risk management, asset pricing, and market simulation, with growing synergy between stochastic calculus and machine learning paradigms.

Classical tools such as the Fokker–Planck equation \cite{van2021control,risken1996fokker} play an important role in modeling the complex behavior of financial returns and currency exchange rate dynamics. A foundational approach was presented by Sornette \cite{sornette2001fokker}, and Renner et al. \cite{renner2001evidence}, who confirmed that high-frequency exchange rate data exhibit Markovian properties and can be effectively described using the Fokker–Planck equation derived from the Kramers–Moyal expansion \cite{nawroth2007medium}. Moreover, Ghasemi et al. showed that the logarithmic return of the price time series is a stationary Markovian process \cite{ghasemi2007markov}. Similarly, Tang et al. \cite{tang2000modelling} and later Lim et al. \cite{lim2007dynamical} examined multiple financial time series, including currency exchange rates, demonstrated that the drift and diffusion coefficients encapsulate essential features of return distributions, volatility clustering, and market shocks. Farahpour et al. \cite{farahpour2007langevin} extended this methodology to emerging markets, developing a Langevin-like equation for exchange rate returns based on empirical estimation of Kramers–Moyal coefficients, showing strong agreement with observed statistical properties. In addition to these, other studies \cite{karth2002stochastic,borland2002theory,michael2003financial,bassler2007nonstationary,popescu2015kramers,anvari2016disentangling} establish a robust precedent for using the Fokker–Planck equation in the modeling of financial systems and motivate its application to volatile currency environments.


In this work, we investigate how the temporal evolution of stochastic properties in exchange rate time series can offer deeper insight into a country's underlying monetary and economic conditions. We focus on the USD exchange rates of Iran, Turkey, and Sri Lanka, three economies that have experienced volatility and policy shifts in recent years, as case studies. Our analysis begins by confirming that the log-return series of these exchange rates exhibit a local stationary and Markovian behavior, thereby justifying the application of a second-order Fokker–Planck equation, as permitted by Pawula’s theorem \cite{Tabar2019}. To capture the dynamic nature of these systems, we employ a rolling window approach \cite{kocc1995analysis}, which applies a fixed-size moving window over the data, to estimate time-varying drift and diffusion coefficients. Using the ruptures Python library, we detect structural breakpoints in these coefficients and observe that abrupt changes often coincide with major political or economic events. These findings suggest that the stochastic dynamics of exchange rate movements is a measurable indicator of macro-level disruptions.

\section{Theoretical and methodological framework}
\label{Theoretical and methodological framework}

We use a methodology to extract stochastic characteristics from empirical time series under the assumption that the time series are realizations of a continuous Markov process. Specifically, we focus on estimating the drift and diffusion coefficients that define the system's underlying dynamics, using the Kramers–Moyal expansion and its connection to the Fokker–Planck and Langevin formalisms.

The time evolution of the probability density function \( p(x,t) \) for a Markovian process is governed by the Kramers–Moyal expansion:
\begin{equation}
\frac{\partial p(x,t)}{\partial t} = \sum_{n=1}^\infty \frac{(-1)^n}{n!} \frac{\partial^n}{\partial x^n} \left[D_n(x,t) p(x,t)\right],
\end{equation}
where $x$ and $t$ represent the random variable and time, respectively. The \( n \)-th Kramers–Moyal coefficient is defined as
\begin{equation}
D_n(x,t) = \lim_{\Delta t \to 0} \frac{1}{\Delta t} \int_{-\infty}^{\infty} (x' - x)^n p(x', t + \Delta t \mid x, t)\, dx'.
\end{equation}
In practice, this expansion is cut off at the second order—a reduction justified by Pawula's theorem when \( D_4(x,t) = 0 \) leading to the Fokker–Planck equation. According to Ito integral \cite{kelly2016smooth}, this yields an equivalent stochastic differential equation of the Langevin type:
\begin{equation}
\frac{dx}{dt} = a(x, t) + b(x, t)\, \Gamma(t),
\end{equation}
where \( \Gamma(t) \) is Gaussian white noise satisfying \( \langle \Gamma(t) \rangle = 0 \) and \( \langle \Gamma(t_1) \Gamma(t_2) \rangle = \delta(t_1 - t_2) \). The drift and diffusion functions are related to the first and second Kramers–Moyal coefficients via
\begin{align}
D_1(x,t) &= a(x,t), \\
D_2(x,t) &= \frac{1}{2} b^2(x,t).
\end{align}

To apply this framework to empirical financial data, we focus on daily exchange rate time series denoted by \( S(t) \), representing the daily price of a foreign currency in domestic units. Since \( S(t) \) is strictly positive by definition, it is common practice to consider its logarithm, \( \log S(t) \), to facilitate mathematical treatment and stabilize variance. However, \( \log S(t) \) typically exhibits nonstationary behavior due to long-term trends or structural shifts. To address this, we analyze the log-returns, defined as
\[
r(t) = \log\left(\frac{S(t+\Delta t)}{S(t)}\right),
\]
which approximate the first differences of the log-price. This transformation yields a locally stationary time series under mild assumptions and is widely used in both theoretical and empirical studies of financial markets \cite{ghasemi2007markov}.

In this study, we focus on three US dollar exchange rate time series in countries that have experienced significant currency volatility and major economic disruptions in recent decades. Specifically, we analyze: (i) Iran, over the period from March 2002 to February 2025 (unofficial US currency exchange market); (ii) Turkey, from January 2010 to May 2025; and (iii) Sri Lanka, from January 2005 to May 2025.
These countries provide representative case studies where external shocks, structural shifts, and policy instability have had measurable impacts on the evolution of exchange rates. As a result, the associated time series exhibit complex stochastic behavior—including potential regime shifts —that are well-suited to the methodological framework developed in this work.


We adopt a method based on the Chapman-Kolmogorov equation to assess the Markovian property of the time series \cite{Tabar2019}. Specifically, we evaluate the quantity \( Q_M(T) \), which measures the deviation between the two-step conditional probability \( p(x_3, t+2T \mid x_1, t) \) and the Chapman-Kolmogorov composition \( \int p(x_3, t+2T \mid x_2, t+T) p(x_2, t+T \mid x_1, t) dx_2 \). The empirical value of \( Q_M(T) \) is then obtained by computing the \( L^1 \) norm of the difference between the directly estimated and the composed transition probabilities. Empirically, we find that \( Q_M(T) \) decays approximately as \( e^{-T/T_M} \), where \( T_M \) is interpreted as the Markov length of the process. We estimated that the Markov time scales for the US dollar exchange rate for Iran, Turkey, and Sri Lanka is approximately 0.78, 0.80, and 0.78 days, respectively (see the Supplementary Material, and Fig. \ref{fig:markovian_analysis}). These results are consistent with findings of other studies demonstrated the Markovian properties of empirical financial time series \cite{Tabar2019,kalekar2024applications,abi2025volterra}.

Assuming local stationarity and ergodicity \cite{ghasemi2007markov}, we estimate the Kramers–Moyal coefficients from a single time series \( X_i=X(t_i) \), sampled at uniform intervals \( t_i = t_0 + i \Delta t_0 \). The empirical estimates are given by:
\begin{equation}
D_n(x) \approx \frac{1}{\Delta t_0} \left\langle (X_{i+1} - X_i)^n \mid X_i \approx x \right\rangle.
\end{equation}
To implement this numerically, the state space of \( X \) is discretized into \( N \) bins of width \( \Delta X = (X_{\max} - X_{\min}) / N \). For each bin centered at \( X_i \), we identify time points where \( X_i \) falls within the bin and calculate the conditional moment:
\begin{equation}
D_n(x) = \frac{1}{\Delta t_0} \left\langle (X_{i+1} - X_i)^n \mid X_i \in \text{bin centered at } x \right\rangle.
\label{D_n}
\end{equation}
The suitable size of the bin in the above equation should be chosen; When the number of bins is too large, many will lack sufficient data points for reliable estimation in Eq. \ref{D_n}. Conversely, too few bins yield overly coarse Kramers–Moyal coefficients, limiting the accuracy and interpretability of the results. This binning procedure enables the estimation of local, state-dependent coefficients, thereby reconstructing the stochastic dynamics from discrete financial time series.

\section{Results}

\subsection{Log-return time series}

\begin{figure}[H]
    \centering

        \centering
        \includegraphics[width=\textwidth]{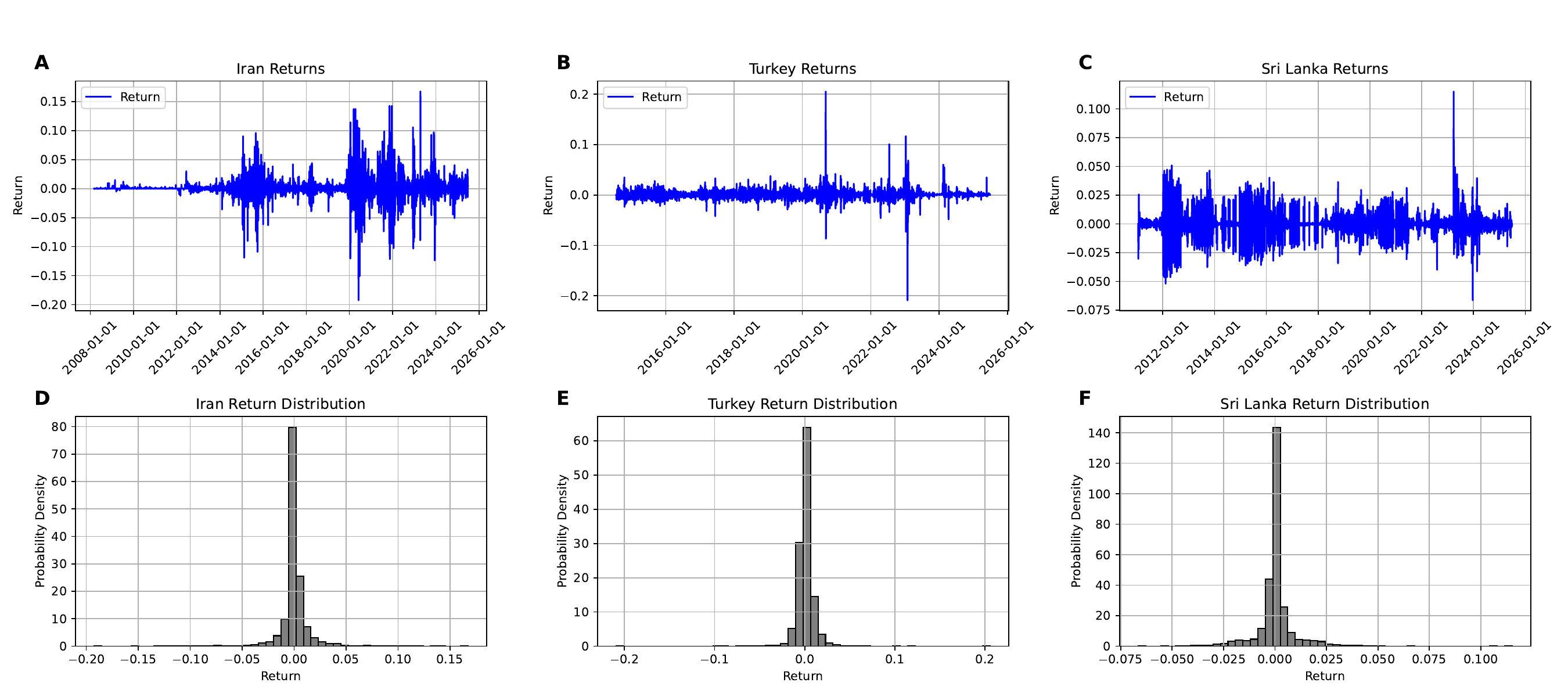}
        \caption{Daily log-returns (A-C) and their empirical probability distributions (D-F) for the US exchange rates of Iran, Turkey, and Sri Lanka. }
    \label{price statistics}

    \hfill
\end{figure}

Fig.~\ref{price statistics} shows the time series of the log-return variable \( r(t) \) for the daily price of one US dollar for Iran (in rials), Turkey (in lira), and Sri Lanka (in rupees), along with its distribution. As can be seen, the returns typically appear in clusters, indicating periods of heightened or reduced volatility. Additionally, the distribution of \( r(t) \) over time (bottom row) shows that the majority of values lying within a narrow range around zero, which aligns with the expectation that exchange rates remain mostly stable with occasional fluctuations. This central concentration also reflects the inherent noise and stochastic nature of the system.

While extreme events are indeed important, their rarity (evident in the tails of the return distribution) makes it difficult to compute reliable averages. Therefore, we restrict further analysis to the domain $|r(t)| \leq 1.5 \sigma$, where $\sigma$, where the data are the most densely populated. 

\subsection{Jump coefficients \texorpdfstring{$D_1$, $D_2$, and $D_4$}{D1, D2, and D4}}

\begin{figure}[H]
    \centering

        \centering
        \includegraphics[width=\textwidth]{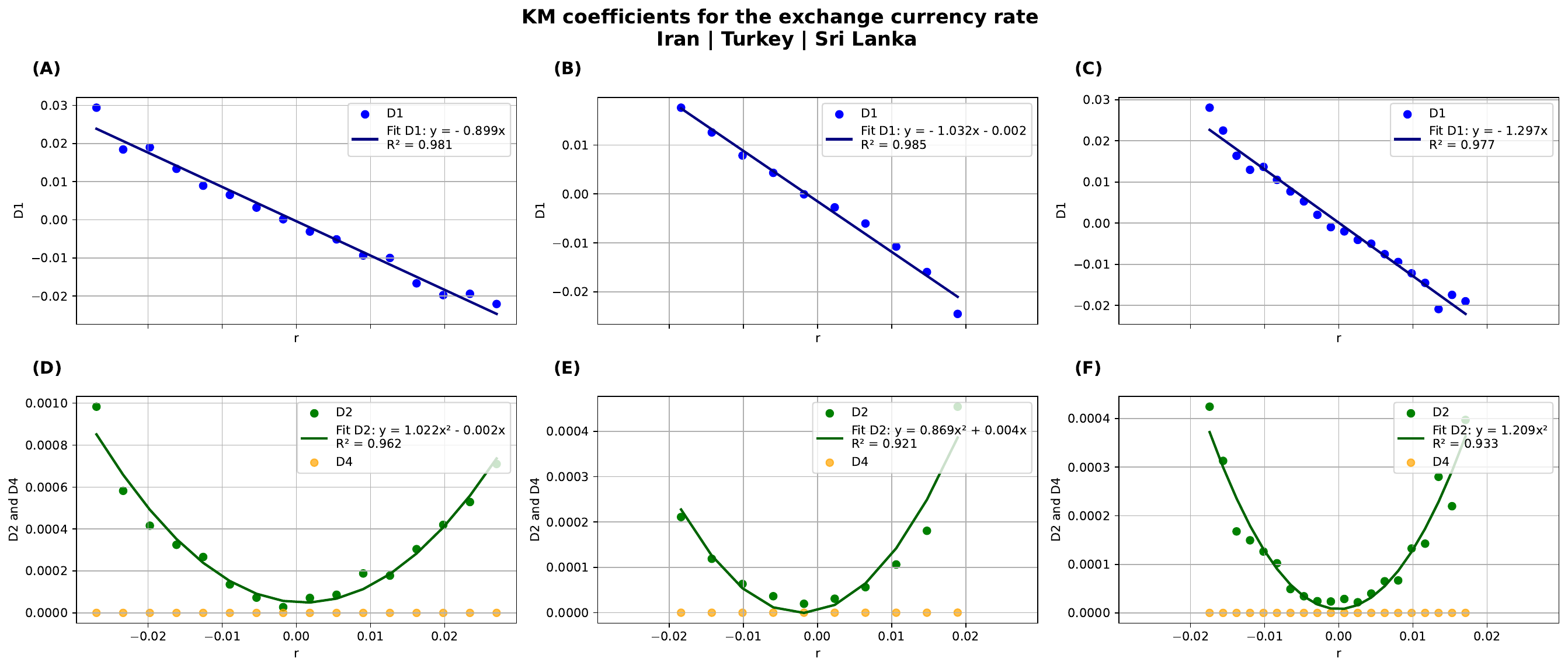}
        \caption{Kramers–Moyal (KM) coefficient analysis of the US exchange rate returns for Iran, Turkey, and Sri Lanka. Top row: Estimated first KM coefficient $D_1(r)$ (drift) with linear fit and $R^2$ goodness-of-fit score. Bottom row: Second KM coefficient $D_2(r)$ (diffusion) with quadratic fit and fourth-order moment $D_4(r)$ over the same bins. The $r$-axis represents the range of return values used for binning.}
        
        \label{km_coefficients_plot}

    \hfill
\end{figure}

Fig.~\ref{km_coefficients_plot} shows the estimated jump coefficients \( D_1(r) \), \( D_2(r) \), and \( D_4(r) \) plotted against the return variable \( r \) for the three countries. Across all cases, the drift coefficient \( D_1(r) \) exhibits a strong linear dependence, consistent with mean-reverting behavior in the exchange rate dynamics. The diffusion coefficient \( D_2(r) \) demonstrates a clear nonlinear dependence on \( r \), captured by a second-order polynomial fit, suggesting state-dependent volatility. This observation is consistent with findings from previous studies that report non-constant and return-dependent volatility in financial time series (see\cite{ghasemi2007markov} for instance).

To assess the validity of the Fokker–Planck approximation, we examine the magnitude of \( D_4(r) \) relative to \( D_2(r) \). In all three data sets, \( D_4(r) \) remains negligible compared to \( D_2(r) \), satisfying the conditions of the Pawula's theorem \cite{pawula1967approximation} and thereby justifying the truncation of the Kramers–Moyal expansion at second order. This supports the use of a Langevin-type stochastic model with state-dependent drift and diffusion terms \cite{Tabar2019}.

\subsection{Rolling window estimation of \texorpdfstring{$D_1$ and $D_2$}{D1 and D2}, and breakpoint distribution}

In addition to visualizing the fitted KM coefficients, we express the drift and diffusion terms using time-dependent parameterizations:
\begin{equation}
D_1(t, r) \approx \alpha(t)\, r + \gamma(t), \qquad D_2(t, r) \approx \beta(t)\, r^2 + \delta(t)\, r 
\end{equation}
Here, \( \alpha(t) \) and \( \beta(t) \) are time-dependent coefficients that characterize the linear and quadratic dependence of the drift and diffusion terms, respectively. These coefficients capture the evolving influence of the underlying stochastic dynamics over time. The additional terms \( \gamma(t) \) and \( \delta(t) \) represent time-varying shifts and curvatures in the system. Analyzing the temporal behavior of these parameters provides insight into changes in system structure, potentially revealing signs of instability or regime transitions.

In particular, the linear form of the drift, with its negative slope and near-zero intercept, suggests a mean-reverting tendency in the dynamics. This behavior may arise from internal market mechanisms or from external interventions by governments or central banks aimed at stabilizing the exchange rate. Meanwhile, the quadratic structure of the diffusion indicates that market noise and unpredictability intensify as the return deviates further from its steady state near zero.

 To investigate the temporal behavior of these coefficients, we implement a rolling window approach. Instead of estimating the Kramers–Moyal coefficients over the entire time series, we divide the data into overlapping segments of fixed length and apply the estimation procedure within each window \cite{kocc1995analysis}. For each segment, we fit the conditional moments \( D_1 \) and \( D_2 \) using linear and quadratic functions of \( r \), respectively, extracting the local values of \( \alpha(t) \) and \( \beta(t) \). By sliding the window across the full time series, we obtain time series of the estimated coefficients \( \alpha(t) \) and \( \beta(t) \).

To ensure reliability in the local regression fits, we impose a constraint that the coefficient of determination (regression score) for each fit must exceed 0.8. Windows that fail to meet this criterion are excluded from the final estimates.

\begin{figure}[H]
    \centering

        \centering
        \includegraphics[width=\textwidth]{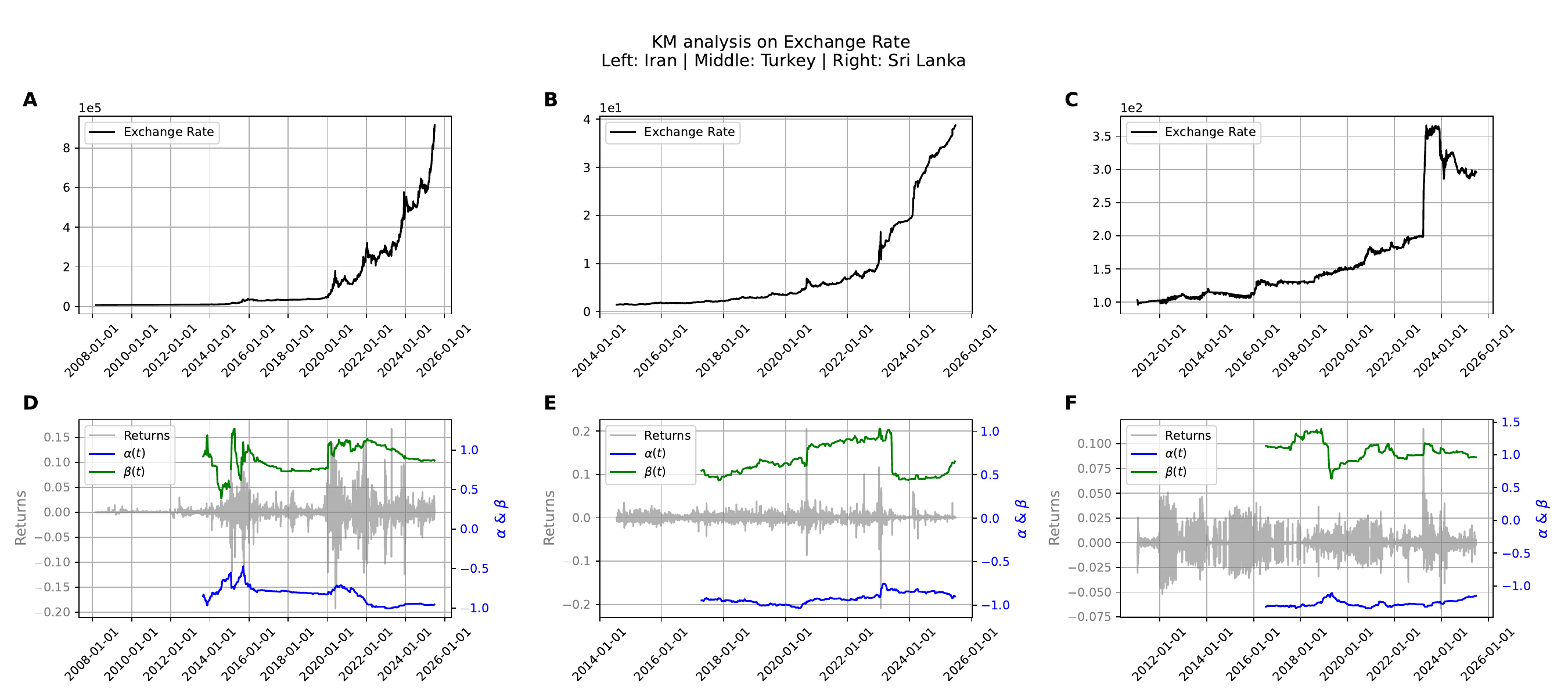}
        \caption{Exchange rate dynamics and KM coefficient estimates for Iran, Turkey, and Sri Lanka. The top row (A-C) shows the exchange rate time series in each country. The bottom row (D-F) presents the daily return series (gray), overlaid with the estimated drift $\alpha(t)$ (blue) and diffusion $\beta(t)$ (green) coefficients computed from the Kramers–Moyal (KM) expansion. Results are based on rolling-window estimation and averaged over a range of bin configurations (see the main text for more information).}

        \label{dinamics_of_d1_d2}

    \hfill
\end{figure}

Fig.~\ref{dinamics_of_d1_d2} shows the results of implementation of the above method for the three countries.  We used a rolling window of 2000 days for Iran and Sri Lanka, and 1000 days for Turkey to account for differences in data length and market behavior. To mitigate the sensitivity of the Kramers--Moyal coefficient estimates to the choice of bin size, we computed \( D_1 \) and \( D_2 \) for multiple bin counts in the range \(30\text{--}100\) with a step of 5, and averaged the results to obtain more stable and robust estimates (See Supplementary Information and Section \ref{Sensitivity Analysis} and Fig.~\ref{fig:sensitivity_analysis}). 

As can be seen in Fig.~\ref{dinamics_of_d1_d2}, the trajectories of \( \alpha(t) \) and \( \beta(t) \) reveal substantial temporal variation, highlighting the presence of instability, structural breaks, and potentially regime-switching behavior in the exchange rate dynamics of each country.

\subsection{Interpretation of breakpoints and temporal instability in currency markets}

Structural breaks in the estimated coefficients \( \alpha(t) \) and \( \beta(t) \) can offer meaningful insights into shifts in the underlying dynamics of currency markets. These shifts may correspond to changes in political regimes, economic policy, or monetary interventions. Theoretically, an increase in \( \alpha(t) \) implies a stronger mean-reverting tendency in the exchange rate dynamics, while an increase in \( \beta(t) \) reflects heightened sensitivity to stochastic fluctuations—potentially indicating elevated market uncertainty or external shocks \cite{dua2021regime}.

In addition, Early Warning Signals (EWS) are statistical indicators that can anticipate critical transitions or sudden regime shifts in dynamical systems \cite{george2023early,lade2012early}. As an example, in some cases, when a system approaches a tipping point, its recovery rate from small perturbations tends to slow down—a phenomenon known as critical slowing down \cite{dakos2008slowing,dai2012generic}. This leads to measurable changes in time series properties, most notably an increase in autocorrelation and variance \cite{bury2021deep}. In the context of financial systems, abrupt rises in these indicators may serve as precursors to major structural changes driven by shifts in policy, external shocks, or systemic instability \cite{diks2019critical,song2024early}. Monitoring such signals in the estimated dynamics can thus provide valuable foresight into emerging transitions before they fully manifest.

By leveraging the time-dependent structure of \( \alpha(t) \) and \( \beta(t) \), we obtain a data-driven measure of market instability. Specifically, we analyze the occurrence of structural breakpoints in these coefficient time series to identify periods of instability. We use the \texttt{ruptures} Python library to detect structural breakpoints in time series data, applying binary segmentation with the \texttt{l2} model. In this context, a breakpoint marks a time at which the statistical properties of the series—specifically the mean—undergo a significant shift. The \texttt{l2} model assumes piecewise-constant mean segments, so each detected breakpoint corresponds to a change in the average level of the time series. To avoid overfitting short-term fluctuations, we apply the detection algorithm with relatively low sensitivity. Breakpoints are identified independently in both \( \alpha(t) \) and \( \beta(t) \), and their union is used to construct a comprehensive temporal map of regime shifts. Full details of \texttt{ruptures} is provided in supplementary materials.

Fig.~\ref{alpha_beta_vs_time} illustrates the results of this analysis. The temporal distribution of breakpoints reveals intervals of elevated breakpoint frequency—indicative of high market instability—and other intervals where the dynamics remain relatively stable. These episodes of increased breakpoint density can be plausibly linked to known political or economic disruptions in each country’s history.

For example in case of Iran, several notable events correspond closely to such periods. The years 2014 and 2015 marked the negotiation phase and eventual signing of the Joint Comprehensive Plan of Action (JCPOA), which raised expectations of sanctions relief and introduced significant shifts in market expectations. This relative optimism was disrupted following the U.S. presidential election in November 2016, as Donald Trump’s administration signaled a more confrontational stance toward Iran. The subsequent U.S. withdrawal from the JCPOA in May 2018 and the implementation of a “maximum pressure” sanctions campaign reintroduced severe economic constraints. Another sharp policy shock occurred in November 2019, when the Iranian government abruptly increased fuel prices by up to 300\%, triggering widespread protests and a cascade of economic uncertainty. These events collectively align with the observed regime shifts in the Iranian exchange rate data. In the case of Turkey, periods of high instability closely align with real-world events, including the onset of the COVID-19 pandemic, the Central Bank of Turkey's decision to cut interest rates in early 2020, and escalating political and geopolitical tensions. In the case of Sri Lanka, the years marked by heightened instability correspond with major national crises, such as the outbreak of the COVID-19 pandemic, the political upheaval and government collapse in 2022, and a drastic decline in foreign exchange reserves.

\begin{figure}[H]
    \centering

        \centering
        \includegraphics[width=\textwidth]{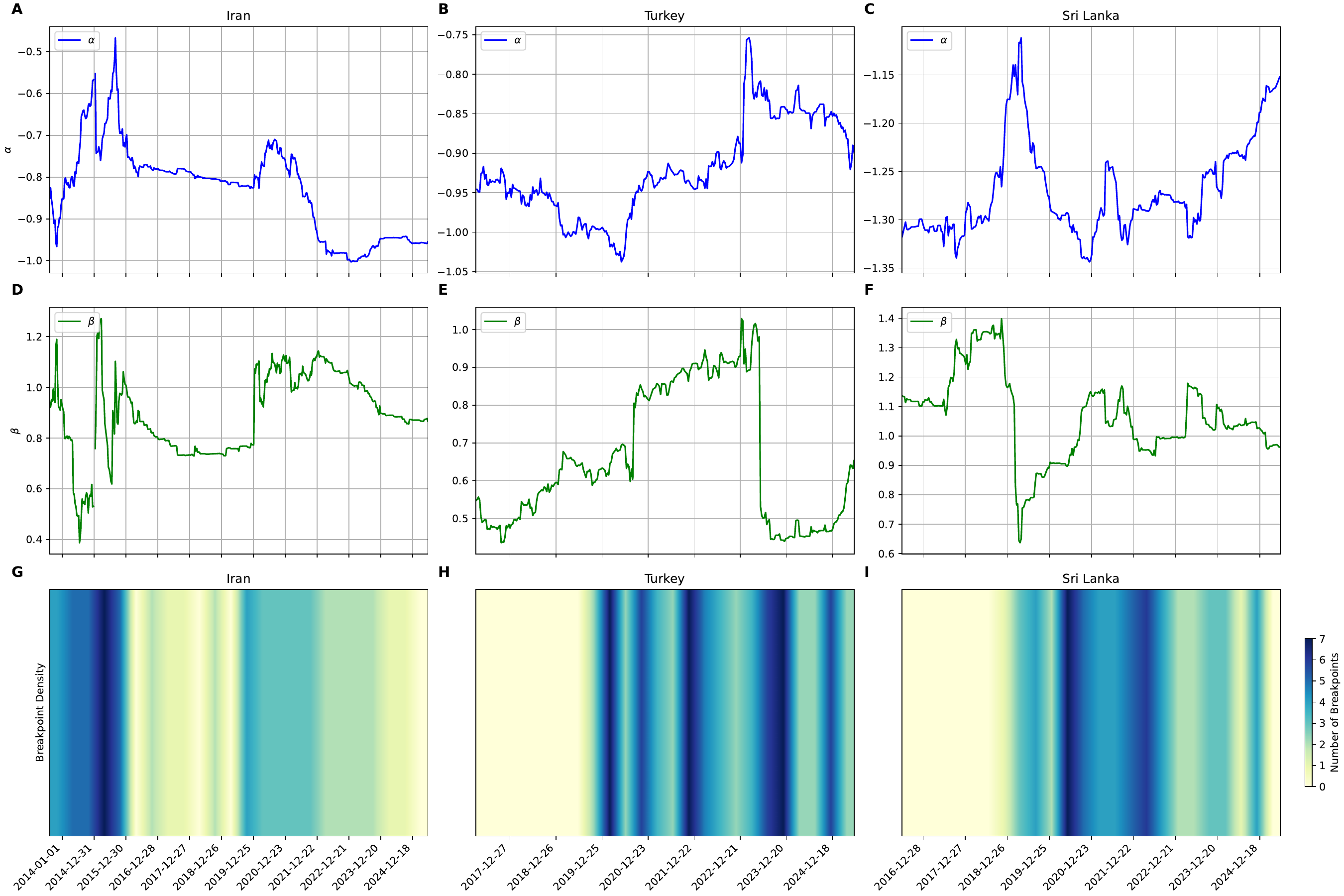}
        \caption{Visualization of $\alpha(t)$ and $\beta(t)$ coefficients (A-C and D-F, respectively), along with corresponding breakpoint density heatmaps (G-F) for Iran, Turkey, and Sri Lanka. Breakpoints are extracted from $\alpha(t)$ and $\beta(t)$ time series using binary segmentation and aggregated into 6-month bins. Higher density regions indicate time intervals with more frequent structural changes in KM dynamics.}

        \label{alpha_beta_vs_time}

    \hfill
\end{figure}

\section{Conclusion}

This study presents a data-driven reconstruction of the stochastic dynamics underlying USD exchange rate time series in volatile economies, specifically Iran, Turkey, and Sri Lanka—using the Kramers–Moyal framework. By empirically estimating the drift and diffusion coefficients from high-frequency return data, we establish that the log-return processes are locally stationary and satisfy the Markov property at daily resolution. This validates the use of a second-order Fokker–Planck equation, and by extension, a Langevin-type stochastic differential equation, to describe the evolution of these exchange rates.

The inferred Langevin coefficients consistently point to a stabilizing drift and a return-dependent (nonlinear) diffusion term across all three countries, suggesting the presence of endogenous stabilizing mechanisms as well as exogenous shocks. Furthermore, our application of a rolling window approach combined with structural breakpoint detection revealed temporal variations in the estimated coefficients that align with key political and economic events. These results underscore the utility of this framework for identifying regime shifts and potential early warning signals in currency markets.

However, several assumptions and methodological constraints limit the generalizability and scope of our findings. First, our analysis was restricted to three case studies, all from emerging or politically unstable economies, limiting extrapolation to more stable currency regimes. Second, due to the need for reliable statistical estimates, we confined our attention to return values within the range $\left| r(t) \right| \leq 1.5\sigma$, excluding rare but significant tail events. This constraint inherently omits subtle non-Gaussian features and extreme fluctuations that may carry important systemic information.

Third, the estimation of drift and diffusion coefficients is sensitive to hyperparameters, particularly the choice of rolling window size and the number of bins used for discretizing state space. Although we performed a sensitivity analysis to mitigate these effects, no universally optimal parameter set exists. Different configurations could produce differing breakpoint patterns or inferred dynamics, indicating a trade-off between resolution and statistical robustness.

Fourth, while the timing of many detected breakpoints coincides with major macroeconomic or geopolitical disruptions, the relationship remains correlational. Causality cannot be established without deeper integration of exogenous variables and counterfactual analysis. Lastly, caution must be exercised in generalizing these findings to stable currency systems or those operating under tightly managed exchange rate regimes, where the nature and magnitude of stochastic dynamics may differ substantially.

Future extensions of this work could explore multi-currency interactions, integrate additional exogenous indicators such as interest rate spreads or trade balances, and incorporate rare-event modeling to address the influence of tail risks. Broadening the methodology across different monetary regimes could further illuminate the diversity of dynamical structures in global currency markets.

\section{Code and Data Availability}

All the codes and empirical data used in this research can be found in the following repository:

\url{https://github.com/Yazdan-Babazadeh/Currents-Beneath-Stability}

\newpage

\bibliographystyle{unsrt}

\newpage

\section{Supplementary information}

\subsection{Assessment of Stationarity}

\begin{figure}[H]
    \centering

        \centering
        \includegraphics[width=\textwidth]{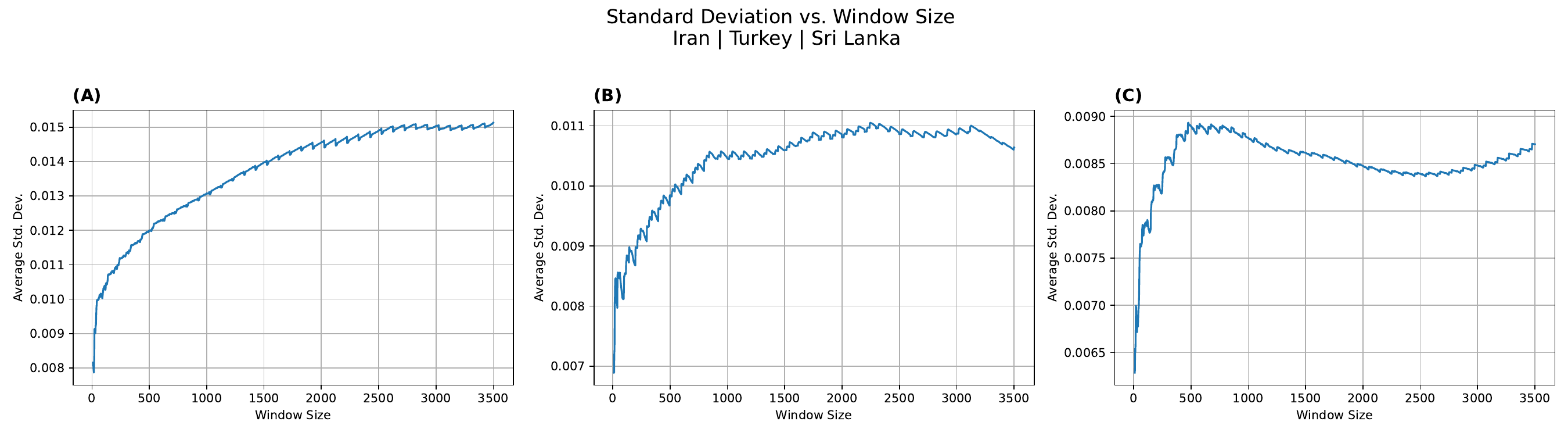}
        \caption{$r(t)$ vs. time. Most values cluster around $r=0$.}

        \label{stationary_analysis}

    \hfill
\end{figure}

To evaluate the stationarity of the log-return process \( r(t) \), we apply a sliding-window variance method. For each window size \( S \), we compute the local variance across the time series and average the results to obtain \( W(S) \). As shown in Fig.~\ref{stationary_analysis}, \( W(S) \) increases initially but quickly saturates, indicating that the variance stabilizes across time. This behavior is consistent with weak stationarity, suggesting that the statistical properties of \( r(t) \) remain stable over time.

\subsection{Mathematics of Markov Processes}

Markov processes are characterized by their memoryless property: the future state of a system depends only on the present state and not on its past history. To formalize this, we consider the price increments $x(\tau)$ over different time scales $\tau$. The key quantities in this framework are the conditional probability density functions. Given a joint distribution $p(x_2, \tau_2; x_1, \tau_1)$ for increments at two nested scales $\tau_1 < \tau_2$, the conditional density is defined as
\begin{equation}
p(x_2, \tau_2 | x_1, \tau_1) = \frac{p(x_2, \tau_2; x_1, \tau_1)}{p(x_1, \tau_1)}.
\end{equation}
Higher-order conditional densities follow a similar construction. A process is Markovian in $\tau$ if the conditional probability of an increment at a finer scale depends only on the increment at the next coarser scale:
\begin{equation}
p(x_N, \tau_N | x_{N-1}, \tau_{N-1}; x_{N-2}, \tau_{N-2}; \dots; x_1, \tau_1) = p(x_N, \tau_N | x_1, \tau_1),
\label{asqar}
\end{equation}
for $\tau_1 < \tau_2 < \dots < \tau_N$. As a result, the full $N$-point joint probability density can be factorized as a product of two-point conditional densities:
\begin{equation}
p(x_N, \tau_N; \dots; x_1, \tau_1) = \prod_{i=1}^{N-1} p(x_{i+1}, \tau_{i+1} | x_{i}, \tau_{i}) \cdot p(x_1, \tau_1).
\end{equation}
This property significantly simplifies the statistical description of the process and serves as the basis for further modeling.

To analyze the stochastic behavior of exchange rate increments across different time scales, we adopt the framework of Markov processes. A stochastic process $x(\tau)$ is said to be Markovian if the conditional probability density, Eq.~\ref{asqar} depends only on the immediately preceding scale: $p(x_N, \tau_N | x_1, \tau_1)$ for $\tau_1 < \tau_2 < \dots < \tau_N$. Under this condition, the process obeys a master equation known as the Kramers–Moyal expansion. If the fourth-order coefficient $D_4$ vanishes, Pawula’s theorem ensures that the expansion cuts off after the second order, resulting in a Fokker–Planck equation:
\begin{equation}
\frac{\partial}{\partial \tau} p(x, \tau) = \left\{ -\frac{\partial}{\partial x} D_1(x, \tau) + \frac{\partial^2}{\partial x^2} D_2(x, \tau) \right\} p(x, \tau),
\end{equation}
where $D_1$ and $D_2$ are the drift and diffusion coefficients, respectively. These coefficients can be empirically estimated from conditional moments of the data. The underlying Langevin equation,
\begin{equation}
\frac{d}{d \tau} x(\tau) = D_1(x, \tau) + \sqrt{D_2(x, \tau)} f(\tau),
\end{equation}
with $f(\tau)$ being a Gaussian white noise.
\subsection{Markovian Properties}

To assess whether the log-return process \( r(t) \) can be approximated as a Markov process, we implement a numerical test based on the Chapman-Kolmogorov equation. Specifically, we estimate the deviation from Markovianity using a function \( Q_M(T) \), which compares direct and composed transition probabilities over a time delay \( T \). 

We divide the time series into overlapping triplets \( (r(t), r(t+T), r(t+2T)) \), and discretize the return space into \( N = 100 \) bins. Using these, we construct three transition probability matrices:
\begin{itemize}
    \item \( P_1(b, a) \): the transition probability from bin \( a \) at time \( t \) to bin \( b \) at time \( t+T \),
    \item \( P_2(c, b) \): from bin \( b \) at time \( t+T \) to bin \( c \) at time \( t+2T \),
    \item \( P_3(a, c) \): the direct transition from \( a \) to \( c \) over time \( 2T \).
\end{itemize}

The quantity \( Q_M(T) \) is computed as:
\[
Q_M(T) = \sum_{i,j} \left| P_3(i,j) - \sum_k P_1(k,i) P_2(j,k) \right|.
\]
This measures how much the empirical process deviates from satisfying the Chapman-Kolmogorov equation. A Markovian process would yield small values of \( Q_M(T) \), ideally approaching zero for sufficiently large \( T \).

To characterize the timescale at which the process becomes approximately Markovian, we fit the resulting \( Q_M(T) \) curve to an exponential decay model:
\[
Q_M(T) = A \, e^{-T / T_M},
\]
where \( A \) is a scale factor and \( T_M \) is the estimated Markov time scale. 

\begin{figure}[H]
    \centering
    \includegraphics[width=\textwidth]{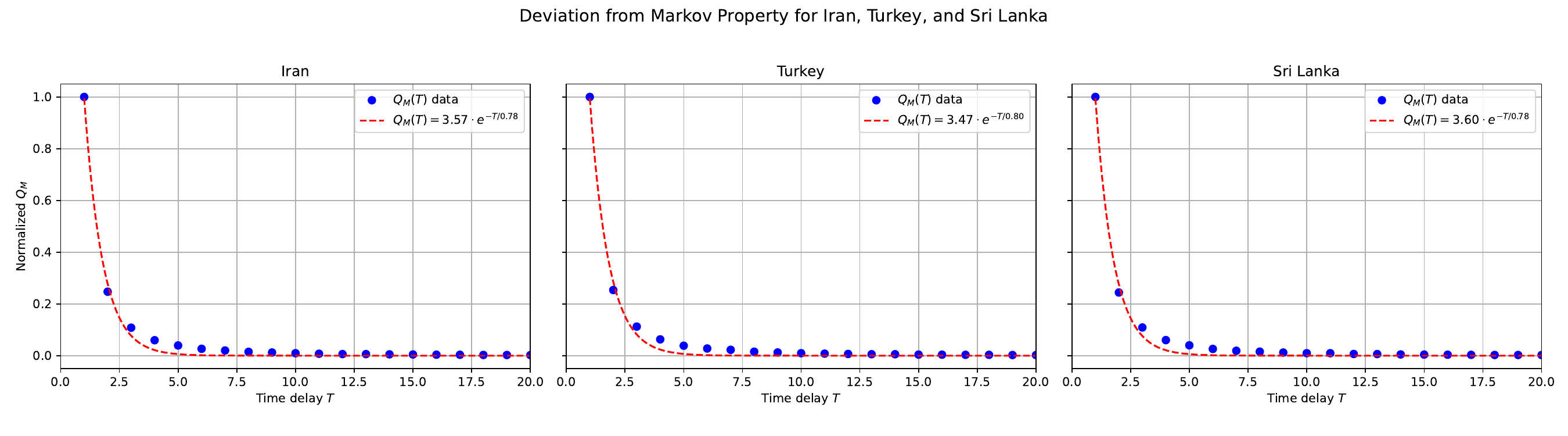}
    \caption{Deviation from Markov Property for Iran, Turkey, and Sri Lanka.}
    \label{fig:markovian_analysis}
\end{figure}

This procedure was applied to the log-return exchange rate return series for Iran, Turkey, and Sri Lanka. Fig.~\ref{fig:markovian_analysis} shows the resulting \( Q_M(T) \) curves and their exponential fits. The estimated values of \( T_M \), in time steps (days), were approximately 0.78 for Iran, 0.80 for Turkey, and 0.78 for Sri Lanka, indicating that the daily log-return processes in all three cases are well-approximated as Markovian at the daily scale. Full implementation details are available in the accompanying code repository.

\subsection{Sensitivity Analysis}
\label{Sensitivity Analysis}

To assess the robustness of the estimated Kramers–Moyal coefficients, we performed a sensitivity analysis of the drift ($\alpha$) and diffusion ($\beta$) terms with respect to variations in the rolling window size. We selected a baseline window size (e.g., 1000 for Turkey) and computed the corresponding time series of $\alpha(t)$ and $\beta(t)$ using a fixed number of bins and step size. We then repeated the estimation for a range of alternative window sizes (e.g., 700 to 1300), interpolating the resulting coefficient series onto the baseline time grid.

\begin{figure}[H]
    \centering
    \includegraphics[width=\textwidth]{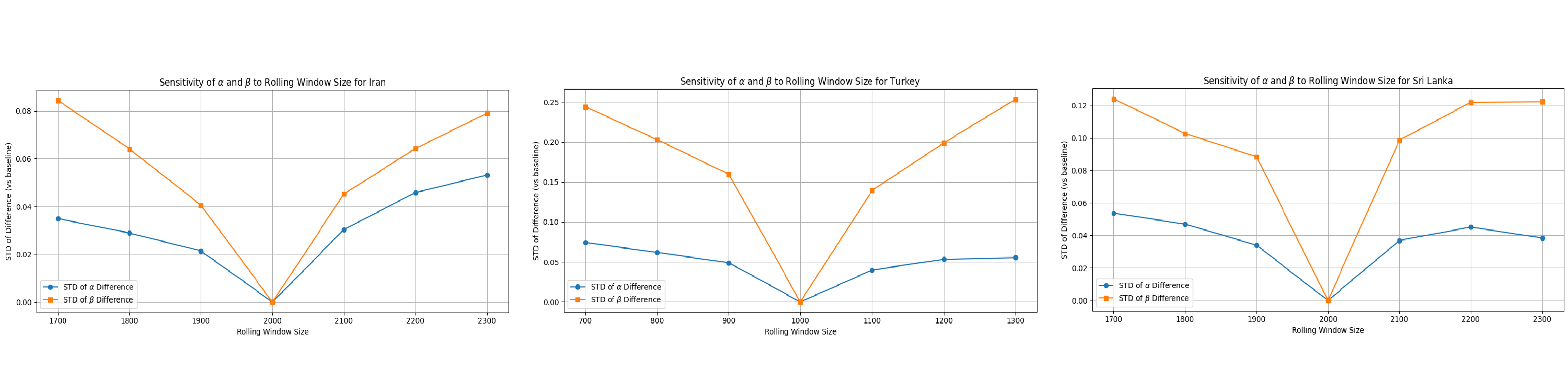}
    \caption{Sensitivity analysis of \( r(t) \) using the sliding-window variance method.}
    \label{fig:sensitivity_analysis}
\end{figure}

For each configuration, we quantified the deviation of the alternative coefficients from the baseline using the standard deviation of the difference at each time point. This approach allowed us to evaluate the absolute sensitivity of the coefficients to the choice of window size. The results, visualized in Figure~\ref{fig:sensitivity_analysis}, reveal how sensitive the drift and diffusion estimates are to temporal aggregation, providing insight into the reliability of the estimated dynamics across scales.

\subsection{Breakpoint Detection via \texttt{ruptures}}

To identify structural changes in the estimated Kramers–Moyal coefficients, we apply change point detection using the \texttt{ruptures} library in Python. This analysis is performed separately on the drift ($D_1$) and diffusion ($D_2$) time series extracted from the stochastic modeling framework.

We employ the Binary Segmentation (BinSeg) algorithm with an $\ell_2$ cost model to detect change points in each time series. This method recursively partitions the signal by identifying the point that most reduces the sum of squared deviations, making it well-suited for detecting abrupt shifts in mean or variance.

The following settings are used in our application of the \texttt{ruptures} framework:
\begin{itemize}
    \item \textbf{Model:} \texttt{"l2"} — specifies that the cost function is based on squared error loss.
    \item \textbf{Algorithm:} \texttt{Binseg} — a greedy, recursive approach for detecting multiple breakpoints.
    \item \textbf{Number of Breakpoints:} $n_{\text{bkps}} = 30$ — the number of change points to identify is fixed to 30 for both $D_1$ and $D_2$ series.
    \item \textbf{NaN Handling:} Missing values in $D_1$ and $D_2$ are imputed using the median of each series via \texttt{np.nanmedian}.
\end{itemize}

The detected breakpoints are used to extract corresponding timestamps, which are subsequently binned into histograms to visualize breakpoint density over time. This enables a comparative analysis of structural regime shifts across countries and time periods. The heatmaps in the bottom row of the figure represent the aggregated density of these breakpoints, revealing periods of heightened volatility or dynamic instability in the underlying exchange rate processes.

\end{document}